\documentclass[twocolumn,showpacs,preprintnumbers,nofootinbib,amsmath,amssymb]{revtex4}

\usepackage{graphicx}
\usepackage{dcolumn}
\usepackage{bm}

\begin{document}

\title{Low energy bounds on Poincar\'{e} violation in causal set
theory}

\author{Nemanja Kaloper}

\author{David Mattingly}

\affiliation{Department of Physics, University of California, Davis,
CA 95616}

\begin{abstract}
In  the causal set approach to quantum gravity, Poincar\'{e} symmetry is
modified by swerving in spacetime, induced by the random lattice
discretization of the space-time structure. The broken
translational symmetry at short distances
is argued to lead to a residual diffusion in momentum space, whereby a particle can
acquire energy and momentum by drift along its mass shell and a
system in equilibrium can spontaneously heat up. We consider bounds
on the rate of momentum space diffusion
coming from astrophysical molecular clouds, nuclear stability and
cosmological neutrino background. We find that the strongest limits
come from relic neutrinos, which we estimate to constrain the momentum space
diffusion constant by $k < 10^{-61} \, {\rm GeV}^3$ for neutrinos
with masses $m_\nu > 0.01 \, {\rm eV}$, improving the previously
quoted bounds by roughly 17 orders of magnitude.
\end{abstract}
\pacs{11.25.-w, 11.25.Mj, 98.80.Cq, 98.80.Qc \hfill astro-ph/0607485}

\maketitle

\section{Introduction}

The construction of a complete theory of quantum gravity has been a
Holy Grail of theoretical physics over the past decades. The quest
for quantum gravity faces new challenges that go beyond the problems
encountered so far when quantizing gauge theories. Theoretically, we
are not even sure how to properly define the observables of quantum
gravity, let alone construct the full dynamics in a
background-independent way. Experimentally, the feebleness of
gravitational force, which may remain weak all the way up to the
$4D$ Planck scale, $m_{\tt Pl} \sim 10^{19}$ GeV, does not give much
hope that direct experimentation could shed more light on the nature
of quantum gravity in the foreseeable future \footnote{An exception
may be that the observed weakness of gravity is just a mirage,
generated by the presence of as yet undiscovered large dimensions
\cite{add}, in which case the full experimentally accessible quantum
gravity might lurk near TeV.}.

Under the looming shadow of conceptual difficulties and without
experimental beacons, one may still pursue a `bottom-up' approach:
one may at least test theoretical proposals for compatibility with
current observations, and constrain the parameters which control new
physics signifying the various short-distance completions. For example, extra
dimensions~\cite{add} or Planck scale modifications of Lorentz
invariance~\cite{Mattingly:2005re}  might both yield experimentally
testable phenomena at low energies
that can differentiate between approaches and lead to useful bounds
on the underlying theoretical frameworks. There may also be other
interesting phenomena, yet to be fully explored.

In this note we will focus on one such novel phenomenon in causal
set theory \cite{blms,sorkin,adgs}, dubbed the `swerve' effect
\cite{DowkerSorkin}. In causal set theory, a particle will hop along
a random lattice, spontaneously acquiring energy-momentum by
drifting along its mass shell, at a rate controlled by a dimensional
parameter that ought to arise from the underlying microscopic
framework. Although this diffusion of particle population's momentum
distribution is Lorentz invariant, the full Poincar\'{e} symmetry
arises only statistically, because the translational symmetries are
violated by the swerves. As a consequence, a gas of particles
will heat up `spontaneously' over time. Clearly the rate of any such
diffusion processes in momentum space must be bounded to avoid
conflicts with observed properties of isolated systems in
equilibrium, which appear largely consistent with energy
conservation.

Here we explore these bounds on the scale of energy-momentum from
the swerves. We consider the possibility that the swerves may 1) heat up old, very
cold molecular clouds in our galaxy, 2) facilitate nuclear
$\alpha$-decay, by providing a channel whereby $\alpha$-particles in
a nucleus gain more kinetic energy and escape more easily, and 3)
render the background relic neutrino density hotter than is allowed
by cosmological bounds on hot dark matter.

We find that molecular cloud heating
and $\alpha$-decay rates yield bounds which are numerically
comparable to the hydrogen heat-up limits of \cite{DowkerSorkin}.
However, the bounds from the background relic neutrino density, that behaves
as hot dark matter, yield much stronger
constraints which we estimate to
restrict the momentum diffusion coefficient to $k \lesssim
10^{-61} \, {\rm GeV}^3$, exceeding those of \cite{DowkerSorkin} by
about 17 orders of magnitude. By noting that different processes
occur at different energy scales, we can also weakly constrain the
dependence of $k$ on energy. In what follows, we briefly review the
basic tenets of causal set theory and the spacetime swerves in
section~\ref{sec:intro}, outline the processes that lead to heating
of nonrelativistic gases in section~\ref{sec:swgas}, and discuss the
bounds in section~\ref{sec:bounds}. We set $\hbar=c=k_B=1$
throughout.

\section{Causal Sets for Pedestrians} \label{sec:intro}

We start with a brief review of causal set theory, for a more
complete discussion see~\cite{Sorkin:2003bx}. A causal set C is a
partially ordered set (a poset), consisting of `points' $x,y,...$
and relations $x \prec y$ which encode causal ordering. For $x \prec
y$ one says that $x$ is to the past of $y$. Further, the ordering
obeys two more rules, $i)$ ~ $x \prec y$, $y \prec z \Rightarrow x
\prec z$ and $ii)$ ~ $x \not \prec x$. The former is transitivity,
establishing links between points: if $y$ is in the past of $z$, and
$x$ in the past of $y$, then $x$ is also in the past of $z$.  The
latter forbids closed causal curves. In general one takes a causal
set to be finite, in the sense that for any ordered pair $\{x,y\}$,
such that $x \prec y$, there are a finite number of points $z$ in
between, such that $x \prec z$ and $z \prec y$.

To relate causal set ideas with low energy physics, we need to
retrieve a geometric structure from the underlying causal set by
some coarse graining. One can do so by reverse-engineering, using
the process of `sprinkling', or a random selection of points from a
manifold $M$, such that the probability of picking any point from a
region depends only on the $4$-volume $V$ of the region. Such sets
of points and inherited causal relations can be thought of as causal
sets $C_M$ which approximate $M$ as random lattices on it. Then the
coarse-grained information about the topology of $M$, its metric,
dimension, etc. can be deduced from operators defined strictly on
$C_M$, and so $C_M$ contains all of the necessary geometric
information at this level of precision.

In contrast to conventional discrete lattices with a fixed lattice
spacing, which explicitly break Lorentz symmetry, it has been argued
that random lattices in causal set theory violate Lorentz symmetry
much more mildly. Since the lattice is random and distributed in the
manifold according to the $4$-volume of spacetime regions, which is
a Lorentz invariant, the symmetry is preserved to a much higher
degree, and should be recovered in the limit when the coarse
graining over the causal set becomes large.

On the other hand, it has been hypothesized by \cite{DowkerSorkin}
that because a causal set is discrete, translation invariance of the
Poincar\'{e} symmetry and hence energy-momentum conservation ought
to be modified. An irregular lattice underlying the macroscopic
spacetime structure won't even preserve a discrete subgroup of
translations, and hence the dual space lattice will also be jagged,
allowing diffusion along a given mass shell instead of an evolution
with a conserved pseudo-momentum, as in the case of a regular
lattice. Indeed, imagine a causal set $C_M$ approximating Minkowski
space, and consider a classical particle initially at rest at a
point $x$ in $C_M$. Its world line is a set of measure zero, and
thus for irregular $C_M$ there won't be any points directly to the
future of $x$. By the arrow of time the particle will move forward
in time, but having no lattice point in the future of $x$ it must
`swerve' slightly to reach the next point in the causal set.
This will endow it with momentum and kinetic energy, which will
therefore not be conserved. The resulting behavior is Brownian drift
in momentum space. For a large population of classical particles at
low energies it should be described by the diffusion equation
\begin{equation} \label{eq:diffusion}
\frac {\partial \rho} {\partial \tau} = k \nabla^2_P \, \rho -
m^{-1} p^\mu \partial_\mu \rho \, .
\end{equation}
Here $\nabla^2_P$ is the invariant Laplacian on the mass shell of
the particle, describing only the drift along mass shell, $m$ is the
mass, $\tau$ is the proper time, $\partial_\mu =
\frac{\partial}{\partial x^\mu}$ is coordinate space derivative and
$\rho$ is the momentum distribution of the population. The constant
$k$ is a diffusion coefficient that sets the rate of diffusion.
Although (\ref{eq:diffusion}) is Poincar\'{e} invariant per se,
given the statistical interpretation of the particle momentum, the
spreading of $\rho$ will yield a change in the average magnitude of
particle momenta at low energies. Thus the standard Poincar\'{e}
symmetry and momentum conservation may be violated statistically.

While the naive `swerve' picture that underlies (\ref{eq:diffusion})
requires modifications at high energies, (\ref{eq:diffusion}) is
singled out as the unique Lorentz invariant diffusion equation for a
population $\rho$, at the lowest order in momentum derivatives.
Indeed, it reflects the condition that a coarse grained description
of a Lorentz-invariant momentum space random motion also respects
Lorentz invariance. Therefore one can argue that (\ref{eq:diffusion})
is a rather generic prediction from causal sets, instead of being
tied to some particular underlying model of particle propagation \cite{DowkerSorkin}.

When energy is bounded below, diffusion will inevitably yield an
increase of average energy and of kinetic energy fluctuations in large
samples of particles. At times short compared to some characteristic
time that controls this classical spreading of the particle energy,
the standard picture based on fully Poincar\'{e} invariant physics
should remain valid. Thus although one doesn't know how to formulate a
full quantum-mechanical description of swerving, which may have to
await the advent of a complete quantum gravity theory based on
causal sets, at short enough time scales one should be able to
separate classical and quantum behavior from each other by virtue of
the Ehrenfest's theorem and simply focus only on the swerve-induced
corrections to the leading-order standard dynamics. The leading
order contribution of swerves to low energy physics phenomena should
appear from the widening of the momentum distribution, inducing a
weak, slow, energy-momentum non-conservation. Over long times, the
aggregate effect of diffusion may interfere with the standard
description and yield observable signatures, and bounds, on the rate
of momentum diffusion.

Such a picture should hold even for agglomerates of particles bound
together, when describing the motion of the center of mass. Indeed,
consider a system of two particles bound together by some potential
interaction. In the momentum space, each particle can be viewed as a
point on mass shell, and the interaction as a `strut' linking them
together. If the free particles would swerve, so should this bound
state, as the underlying nature of spacetime is not affected by the
interaction. However, the problem of finding the allowed lattice
points to which the particles can jump under time evolution will be
complicated because the particles need to jump close enough in order
not to rip up the strut linking them. Sometimes, this may turn out
to be impossible, and the swerves may break up the bond. When the
momentum lattice is dense, with a mean energy spacing of lattice points
much  smaller than the scale of the bond, such destructive jumps will
be very rare, while in the description of non-destructive jumps, whose
particle energy differential due to swerving is smaller than the
binding energy, the resulting picture should be similar to what one expects
for a single elementary state.

The momentum diffusion parameter $k$ has dimensions of energy cubed,
so one would naturally expect it to be some convolution of the
energy scale of the problem and some short distance cutoff that
emerges from causal set theory. To define this scale, let $V_C$ be
the $4$-volume for which on average there is one point inside, and
define the `fundamental' energy scale $E_C= V_C^{-1/4}$. This scale
is the natural analogue of the conventional field theory UV cutoff,
characterizing the momentum space dual lattice, and may be high as
the $4D$ Planck scale $m_{\tt Pl}$. Thus we expect that $k= c \,
{\varepsilon}^{3-\gamma} E_C^{\gamma}$ is a reasonable parameterization
of the diffusion coefficient, where $c$ is
a dimensionless number and ${\varepsilon}$ is the mass
scale of the problem, such that $p^2 = {\varepsilon}^2$. Hence
$\varepsilon$ is invariant on the mass shell, behaving as
the effective mass, in agreement with the microscopic Poincar\'{e}
invariance of (\ref{eq:diffusion}).
Further, in an effective field theory (EFT) framework one might normally expect
that $c \sim {\cal O}(1)$, unless it is protected
by some softly broken symmetry.
In causal set theory, the broken translational symmetries may
be protecting the smallness of $c$, if there is a limit in which they are
restored and swerving completely extinguished. In such a case
$c$ might turn out to be smaller than the naive arguments suggest.
One may wonder if it is even possible to design an EFT framework for swerving,
but as long as swerving is very small, there should exist an approximate description
that may befit. As far as we can tell,
proving this is presently an open question, which we will not pursue here.
Now, if $\gamma$ were large and
negative, $k$ would have rapidly vanished when ${\varepsilon} \ll
E_C$ and so there would have been no diffusion to speak of. On the
other hand, the variance of a population from Brownian motion is a
function of the number of steps taken. If a particle travels a
distance $L$ then as $E_C$ increases the number of steps increases,
allowing for more diffusion. Thus one expects $\gamma$ to be
non-negative\footnote{If the usual dimensional analysis applies to
the determination of $k$, one would expect $\gamma=3$. Keeping $\gamma$
arbitrary we allow for possible anomalous quantum effects.},
possibly allowing interesting diffusion phenomena as time goes on.
In fact, if there existed some kind of effective field theory
description of swerves, one would have thought that the most likely
value of $\gamma$ were three, because this would correspond to describing
the violation of translational symmetry by a marginal operator,
which gives the strongest effects at low energies,
that still respects decoupling.

With this in mind, we now turn to the quantitative constraints on
$k$. We will derive various bounds on $k$ from the temperature
history of astrophysical molecular clouds, $\alpha$-decay, and most
importantly the cosmic neutrino background, which will enable us to
constrain both the absolute value of $k$ and, to a degree, the power
$\gamma$.

\section{Swerving gases}
\label{sec:swgas}

We will be interested in non-relativistic particles, since for very
relativistic ones the  shift of the mean energy of the distribution becomes strongly
suppressed by the boost factors and slows down \cite{broad}.
This can be  glimpsed at by multiplying
(\ref{eq:diffusion}) by a power of $m$, assuming continuity of
$\rho$ and taking the limit $m \rightarrow 0$.
This leaves $p^\mu \partial_\mu \rho \rightarrow 0$, which for a homogeneous and
isotropic distribution yields $\partial_t \rho \rightarrow 0$. On the
other hand, for a non-relativistic population the proper time $\tau$
reduces to the coordinate time $t$, and the evolution equation
(\ref{eq:diffusion}) reduces to \cite{DowkerSorkin}
\begin{equation} \label{eq:diffusiont}
\frac{\partial \rho} {\partial t} = k \nabla^2_P \, \rho \, ,
\end{equation}
where $\nabla^2_P$ is now the standard spatial Laplacian operator on
momentum space $\mathbb{R}^3$. This follows because the
non-relativistic limit $m \gg p$ really corresponds to the
restriction of the dynamics to a region about the mass shell minimum
of size $p \ll m$, where the mass shell degenerates to a $3D$
Euclidean momentum space.

The general causal solution to (\ref{eq:diffusiont}) can be derived
using standard Green's function techniques~\cite{Andersson}.  The
time dependent solution for $\rho(p,t)$ with initial conditions
$\rho(p,0)=f(p)$ is given by
\begin{eqnarray} \label{eq:generalsol}
\rho(p,t) &=& \int_{-\infty}^\infty G(p,p',t,0) f(p') d^3p'
\nonumber
\\ &=& \int_{-\infty}^\infty (4\pi k t)^{-3/2} e^{- \frac {(p-p')^2}
{4kt}} f(p') d^3 p' \, ,
\end{eqnarray}
where $G(p,p',t,0)$ is the Green's function for the diffusion
equation.   One can immediately see the heating effect of swerves in
a population by considering a collection of particles all initially
at rest, $f(p')=\delta^3(p')$. In this case the integrals over $d^3
p'$ can immediately be done and one has simply
\begin{equation}
\rho(p)=(4\pi k t)^{-3/2} e^{- \frac {p^2} {4kt}} \, ,
\end{equation}
which is the thermal distribution for a non-relativistic gas at
temperature $T=2kt/m$, as mentioned in~\cite{DowkerSorkin}.

However in what follows we will consider the initial distribution of
a population which is not at rest but is a ``gas'' at an initial temperature
$T_0$. This is consistent with the picture where the heating
up from swerves is slow compared to the processes governed by
conventional physics. Thus we need
a solution that describes a system that starts with the standard
Maxwell-Boltzman distribution $f(p')= (2 \pi m T_0)^{-3/2}
e^{-E'/T_0}$, and evolves by swerving into
\begin{equation}
\rho(p,t)= \int_{-\infty}^\infty (8 \pi ^2 k t m T_0)^{-3/2} e^{-
\frac {(p-p')^2} {4kt}} e^{- \frac {p'^2} {2 m T_0}} d^3p' \, .
\end{equation}
Factoring the constant part out and reducing the remaining integral
to a Gaussian by completing the square in the exponent, one finds
\begin{equation}\label{eq:heatupmain}
\rho(p,t)= (2 \pi m T)^{-3/2} e^{-p^2/(2 m T)} \, ,
\end{equation}
where $T=T_0 + 2kt/m$.  Hence a gas initially in thermal equilibrium
will retain the same distribution under the influence of swerves.
The temperature, however, increases linearly with time according to
\begin{equation} \label{eq:dTdt}
\frac {dT} {dt}=\frac {2k} {m} \, .
\end{equation}
Having thus established the primary effect of swerves in
(\ref{eq:heatupmain}), (\ref{eq:dTdt}), we can turn to exploring the
bounds on $k$.

\section{Bounds} \label{sec:bounds}

\subsection{Nonrelativistic gases}

From the nature of the swerve effect, one may expect that the
strongest bounds will arise from the consideration of old and cold
systems of light particles. Indeed, the bounds arise from limiting
the gain of mean kinetic energy $\Delta \langle E \rangle \sim
\Delta T$ due to swerves, which from (\ref{eq:dTdt}) constrains $k$
according to $k \lesssim m \Delta E/t$. Heavier particles will gain
less energy due to their inertia, and clearly if the systems haven't
existed for very long or if the energy thresholds are high, the
bounds will be weak. In \cite{DowkerSorkin}, the authors pursued
this reasoning and limiting the rate of increase of temperature of a
very cold hydrogen gas in normal laboratory conditions due to
swerving by a millionth of a degree per second, they found the bound
$k \lesssim 10^{-44} \, {\rm GeV}^3$.

One might hope to get better bounds from very old and cold
astrophysical molecular clouds, which may be sensitive to
heating by swerves. For example, in our galaxy there is
a large molecular cloud called the Edge Cloud 2~\cite{Carnegie}
(EC2), at roughly 28 kpc from the galaxy center. It is known to
contain several species of non-relativistic light molecules,
including e.g. $NH_3$, $CO$, and while it may even contain lighter
molecules such as $H_2$, to be conservative we will use here only
those molecules which have been explicitly confirmed. Naively, this
could yield a stronger bound\footnote{Given that molecular clouds
are not in exact thermal equilibrium and can even have strong
localized deviations, these results must be considered
approximate.} than lab heating limits of \cite{DowkerSorkin}:
limiting the temperature gain due to swerves by the average kinetic
temperature of the $NH_3$ molecules in EC2, $T \simeq 20$ Kelvin,
taking its estimated age as comparable to the age of the Milky Way, and
using the molecular mass of approximately 17 GeV for $NH_3$ and
finally and {\it crucially} assuming no dissipation, we would find a
bound on $k$ from (\ref{eq:dTdt}) of $k\lesssim 10^{-51} \, {\rm
GeV}^3$ that would appear to improve the hypothesized bound in
\cite{DowkerSorkin} by about seven orders of magnitude.

However the assumption that dissipation is negligible is incorrect.
The heat losses from electromagnetic emission reduce the above bound
dramatically. The point is that since the usual molecular clouds are
composed of bound states of charges, they will behave as a system of
elementary dipoles and emit electromagnetic waves when heated. Since
the rate of energy gain from swerves is exceedingly small, and the
clouds are not ultra-cold, the energy losses from electromagnetic
radiation are faster and can neutralize the energy gain from swerves
quickly, by a slight shift of the equilibrium temperature. Due to
the uncertainties in determining this temperature, the quantitative
bounds on $k$ are weakened. To see this, let us consider a gas in
equilibrium with its environment at a temperature $T$, determined to
a small uncertainty $\Delta T \ll T$. At lowest order the
uncertainty in the rate of energy loss due to radiation for a
blackbody of surface area $A$, ignoring e.g. the temperature of the
environment, such as CMB, with $c = \hbar = k_B = 1$ is
\begin{equation}
\Delta \frac {dE} {dt} \simeq \frac{4 \pi^2}{60} A T^3 \Delta T \, .
\end{equation}
Approximating $\frac{4 \pi^2}{60} \sim 1$, and balancing this
against the energy gain from swerves (\ref{eq:dTdt}), $dE/dT= 3 N
k/m$, where $N$ is the number of particles, yields $A T^3 \Delta T =
\frac {3 N k} {m}$. For a roughly spherical system, the lower limit
on $k$ that can be achieved by this is therefore
\begin{equation} \label{eq:limitk}
k \lesssim \frac{m T^3 \Delta T}{n R} \, ,
\end{equation}
where $n$ is the number density and $R$ is the linear dimension of
the system.  Using $n \sim 10^3$ particles per cubic centimeter
\cite{Carnegie}, and a size on the order of a $R\sim $kpc for EC2
the best limit we can achieve on $k$ due to radiative emission is
then
\begin{equation}
k \lesssim 10^{-41} \frac {\Delta T} {1 ~{\rm Kelvin}}  \, {\rm
GeV}^3.
\end{equation}
The uncertainties in temperature for EC2 are within an order of
magnitude of $1$ Kelvin and so the limits on $k$ really are $k \leq
10^{-41} \, {\rm GeV}^3$. In reality, this bound will be better when
we properly include the grey body factors due to the material nature
of emitting gases, that would obstruct the emission of
electromagnetic waves, improving the bound by, at most, a couple of
orders of magnitude and making it marginally comparable with the
bounds claimed in \cite{DowkerSorkin}. We won't pursue those details
here explicitly, because in what follows we will find a considerably
stronger bound from the cosmological relic neutrino background.

This discussion however reveals that the problem of radiative losses
will plague most attempts to constrain swerves. For example, one
might try to use the atmospheres of Sun and Earth, which are
composed of light gases and have been around for about five billion
years \footnote{According, at least, to scientific sources.}, and
whose mean temperatures should not have varied by more than, say,
${\cal O}(1)$. However as Sun and Earth equilibrate by radiative energy
transfer, the bounds on $k$ that one could get from constraining
atmospheric temperature variation do not improve tremendously the
numbers we quoted so far. Some of the problems with radiative losses
could be mitigated by considering ultra-cold materials, for which
the emission of electromagnetic waves, proportional to $T^4$, will
be dramatically reduced.  For example, atom chip Bose-Einstein
condensate (BEC) experiments with
87-Rubidium have been able to create a BEC with temperatures below a
$\mu {\rm Kelvin}$ that contain 2000 atoms in a volume $V \approx
10^{-3}$ mm$^3$ \cite{BEC}. Using these parameters we find that we
can neglect radiative losses if the diffusion parameter $k$ is
larger than $k \sim 10^{-54} \, {\rm GeV}^3$. On the other hand,
the actual measured heating rate of these systems is $0.5~\mu {\rm
Kelvin}/sec$, giving a bound on $k$
\begin{equation}
k \lesssim 10^{-42} \, {\rm GeV}^3 \, ,
\end{equation}
again within two orders of magnitude to the bound claimed in
\cite{DowkerSorkin}. It may be of interest to carefully consider
other such laboratory sources of bounds, however that is beyond the
scope of the present work.

\subsection{Atomic nuclei}

Another example of very long-lived old systems of particles are
atomic nuclei, and those which are particularly useful for our
purposes here are the ones that undergo $\alpha$-decay. As Gamow
showed in 1928 \cite{gamow}, they may be viewed as systems comprised
of a large number of $\alpha$-particles which are bouncing around in
the nuclear potential well. Their average energy controls the
probability of tunneling out of the well, which causes the nucleus
to decay. The whole process is well described by the stationary
limit, where a particle of some typical kinetic energy $\sim {\rm
MeV}$ bounces many times against the barrier walls, occasionally
managing to slip through it. If however an $\alpha$ particle managed
to gain kinetic energy, it would move up inside the barrier,
encountering thinner walls which would be easier to penetrate.
Ultimately, a particle could even climb over the top of the wall and
slip out unhindered.

In the standard Poincar\'e-invariant formulation of quantum
mechanics, energy conservation implies that the energy levels of a
bound state are discrete, as opposed to continuous as in a gas of
free particles. One might therefore think that the
$\alpha$-particles in a nucleus should have discrete energy levels,
and that the gap between them would prevent the swerving-induced
energy gain of any individual particle, instead leading to only an
overall gain of the center of mass energy of the whole nucleus.
However, this intuition is misleading, because in causal set theory
energy is not conserved absolutely.
As a result, one would not get a system with fixed discrete levels,
but with time-dependent energies which would `wobble' around some
statistical mean value, broadening the levels in a way similar to
what happens in many body systems at finite temperature. As a
result, one would expect that a quantum state would really be
represented by energy bands rather than sharp, discrete energy
levels. Indeed, consider for example an $\alpha$-particle in the
ground eigenstate $E_0$ of a static unperturbed Hamiltonian $H$,
where by unperturbed we mean without swerves. If we now include a
small amount of swerving then under time evolution the particle
cannot remain in $E_0$ since time translation is broken by the
swerves. Since the swerves are random the evolution operator must
have non-zero transition probabilities between states as there is
certainly some probability for such a transition to happen all at
once. Therefore the particles must evolve under swerves to a
superposition of states with support on various eigenstates $E_m$ of
$H$, where $E_m>E_0$.  The \textit{mean value} of the
$\alpha$-particles would continuously gain energy, and we can
estimate the rate of this energy gain by its classical value,
appealing to Ehrenfest's theorem.  This should be a good
approximation for deriving bounds on tunneling rates, by essentially
saying that swerving enables tunneling when the broadening of
individual energy levels is comparable to the gap between them in
the Poincar\'e-invariant limit. This justifies the classical
approximation in estimating energy gains in what follows.


Now, a typical stable heavy nucleus has a
lifetime comparable to the age of the universe, $10^{10}$ years, and
there are even some isotopes which are extremely long lived, such as
Bi 209 whose lifetime is about $10^{19}$ years. On the other hand,
the heavy nuclei which are present in nature are believed to have
been created in first stars, and hence they are also very old, with
their current ages comparable to their lifetimes. Because their
measured lifetimes are in very good agreement with the
$\alpha$-decay theory based on the standard Poincar\'{e} invariant
dynamics, we conclude that any possible enhancements of the decay
due to swerving must be within the error of the measured lifetimes.
This is why the longevity of Bi 209 does not help very much: since
it is made artificially, it hasn't been around for long enough to
have swerves destabilize it.

From the functional dependence of the decay rate on the kinetic
energy \cite{blatt},
\begin{equation}
\Gamma \simeq \frac{{\cal A} E {\cal R}}{(E{\cal R})^2 + {\cal B}}
\, , \label{decay}
\end{equation}
where ${\cal R}$ is the effective size of the nucleus and ${\cal
A}$, ${\cal B}$ constants that depend on the barrier properties, the
standard description will remain unaffected by swerves if the total
energy gain of an $\alpha$-particle inside the barrier remains
smaller than a typical kinetic energy of an escaping
$\alpha$-particle. Roughly, this implies that we ought to constrain
$\Delta E \lesssim {\rm MeV}$ over the age of a nucleus, comparable
to the age of the universe. That yields $k \lesssim m_\alpha {\rm
MeV} H_0$, and so we find
\begin{equation}
k_\alpha \lesssim 10^{-44} \, {\rm GeV}^3 \, , \label{alphak}
\end{equation}
which, quite coincidentally, is again comparable with the hydrogen
heating bound of \cite{DowkerSorkin}. Basically, for $k$ smaller
than $k_\alpha$, the diffusion in momentum space is too slow to help
push the $\alpha$-particle out of the nucleus. One might have hoped
to gain a stronger bound from more stable isotopes. However, as we
have noted above, being made in a lab, these are much younger than
the age of the universe, and the swerves could not have had enough
opportunity to build up a large $\Delta E$ yet.

\subsection{Cosmic neutrinos}

We have seen above that the most serious practical obstruction to
constraining the swerve effect arises from radiative energy losses,
that can overshadow the energy gain from swerves already at very low
temperatures. This problem can be overcome if the thermal gas is
composed of particles whose electromagentic interactions are
extremely suppressed. A perfect candidate for such a material are
neutrinos. They are electrically neutral, and while they can have
anomalous electromagnetic dipoles, generated by the quantum loops
involving a lepton and a $W$, the relevant form-factors are very
small. The conservative bounds on the magnetic dipole moment as
quoted by the Particle Data Group \cite{pdg}\footnote{There may be
even stronger bounds from astrophysical and cosmological
considerations, some dating back to \cite{feinberg}, and more in
recent works  \cite{fuku,raffelt}. } restrict the largest magnetic
dipole moment, for $\nu_\tau$, by $\mu < 3.9 \times 10^{-7} \mu_B$,
where $\mu_B$ is the Bohr magneton, and the electric dipole moment,
again of $\nu_\tau$, by $d < 5.2 \times 10^{-17} e {\rm cm}$. For
comparison, these are at least seven orders of magnitude smaller
that the corresponding dipole moments for a hydrogen atom. Because
dipole power losses are proportional to the square of the dipole
moment, the emissivity of the neutrino gas will be at least fourteen
orders of magnitude more suppressed than the emissivity of the
hydrogen gas.

Hence we can treat cold neutrinos as an essentially non-radiating
gas, which can therefore be a sensitive probe of any
Poincar\'e-violating effects that can lead to spontaneous heat up.
In what follows, we will consider implications of these observations
for the cosmological neutrino background. We will show that swerving
can generate enough heat in the cosmic neutrino background such that
it would behave as hot dark matter and conflict with the
observational evidence that dark matter should predominantly be
cold. Requiring this does not happen will yield the strongest bound
on the momentum space diffusion constant, constraining it to roughly
$k \lesssim 10^{-61} {\rm GeV}^3$.

To start, we note that while the cosmological neutrino background
has not been observed
directly, given the successes of the Standard Model of particle
physics and the Big Bang nucleosynthesis, there is
little doubt that it exists (see, for example, a recent review
articles \cite{dolgov,gelmini}). Furthermore, by the observed
neutrino mass differences we believe that two neutrino species, at
least, are heavier than about $0.01\, {\rm eV}$, with at least one species
being heavier than $0.03 \, {\rm eV}$ \cite{whitenu}, while the third may
be lighter, with a temperature slightly below the CMB temperature,
$T_\nu \sim 10^{-4} \, {\rm eV}$. There may of course be even more
light neutrinos, provided that they are Standard Model singlets. If they had masses
in the $10^{-2} \, {\rm eV}$, they would strengthen our bounds even more.

These neutrinos were all in thermal equilibrium with the baryons and
photons in the early universe, at temperatures well above neutrino
masses. Because at those times the neutrinos were relativistic, the
swerve effect was negligible for them, as we have pointed out above.
As the universe expanded, the neutrino gas cooled and heavier
neutrinos became nonrelativistic, so that eventually the swerving
could reheat them again. Radiative energy losses could not
compensate for this energy gain unless the diffusion parameter $k$
were extremely small. To estimate this critical value of momentum
diffusion coefficient $k_{cr}$ induced by swerving, below which
radiative losses become important, we can use a version of Eq.
(\ref{eq:limitk}), with the emissivity factor $\xi < 10^{-14}$
included. Assuming that the heavier neutrino species have masses
$m_\nu > 0.01 \, {\rm eV}$ \cite{gelmini}, are distributed
homogeneously inside the volume of the present horizon size $R \sim
1/H_0 \sim (10^{-33} \, {\rm eV})^{-1}$, with number density $n \sim
T^3_\nu \sim 10^{-12} \, {\rm eV}^3$ and uncertainty  $\Delta T \sim
T_\nu$, to be conservative, we find that the critical value of
$k_{cr}$, below which radiative energy losses become important, is
\begin{equation}
k_{cr} \simeq \xi m_\nu T_\nu H_0 \simeq 10^{-80} \, {\rm GeV}^3 \, .
\end{equation}
If we were to consider these same neutrinos at the time when they
froze out, at $T_\nu \sim m_\nu$, when $H_\nu \sim
(\frac{T_\nu}{T_0})^{3/2} H_0 \sim 1000 H_0$, we could raise
$k_{cr}$ by at most 5 orders of magnitude, which however remains
negligibly small for our purposes here, as will be clear from the
bound we are about to find. Thus, for all values of $k > k_{cr}$ we
can completely ignore the electromagnetic emissions, and treat the
background relic neutrinos as essentially an isolated system.

To understand what happens with the energy density of neutrinos
with the swerving contributions included, in principle we would have
to define, and solve, a cosmological version of the diffusion
equation (\ref{eq:diffusion}),
that would also include the effects of cosmic
expansion on the neutrino distribution and its mean energy. However,
even without tackling this complicated problem, we can understand
the qualitative features of the solution rather simply. Consider a
thermal distribution of particles with number density $n \sim 1/a^3$
and mean energy per particle ${\cal E} \simeq m_\nu + T = m_\nu +
T_0 /a$, where we normalize $a_0 = 1$. Here, $m$ is the rest mass of
the particle species, whereas $T$ is the mean kinetic energy per
particle, or temperature, of the distribution. In the early
universe, $T \gg m_\nu$, but as the universe expands $T$ redshifts
as $1/a$ and eventually becomes negligible compared to $m_\nu$.
Through this epoch, the energy density $\rho_\nu \sim {\cal E} n$
first scales as $\rho_\nu \sim T/a^3 \sim T_0 /a^4$, eventually
changing to $\rho_\nu \sim m_\nu/a^3$.

Enter the swerves. For a typical massive neutrino from the
distribution, the swerves begin to restore its kinetic energy at the
rate $\partial T/\partial t \sim k/m_\nu    $, as per Eq.
(\ref{eq:dTdt}). The swerving slows
after the kinetic energy becomes relativistic again,
reaching $T_* \agt {\rm few} \times m$, and slowly growing on.
While the neutrinos are nonrelativistic,
the mean energy gain per unit time is
approximately constant, so we can estimate the time scale from
$\partial T/\partial t \sim T_*/t_*$ and previous formulas, to be
$t_* \sim m^2_\nu/k$. Thus after $t_*$, while the neutrino number
density continues to dilute as $n \sim 1/a^3$, their characteristic
energy per particle becomes relativistic again,
${\cal E} \agt {\rm few} \times m_\nu$. For
a large enough $k$ this energy continues to swerve up more slowly
thereafter because the time scale of dilution by cosmic expansion is
$\sim H^{-1}$, getting longer as the universe cools, so the
swerving dominates, albeit it slows after particles regain relativistic speeds.
The critical value of $k$ for which enough energy is dumped back
into the neutrinos to compensate for cosmic cooling
can be easily approximated by
equating the cooling and heating time scales ($H^{-1} \sim t_*$ in
this case).  As long as $k$ is above this value then on average we
will end up with a distribution of relativistic particles whose
energy density dilutes only as $\rho_\nu \agt {\rm few} \times
m_\nu/a^3$, instead of the usual $\rho \sim 1/a^4$ law.

Because massive neutrinos started out in equilibrium with other
relativistic species in the universe, such as CMB, after $t_*$ we
can approximate their residual energy density per species as
$\rho \agt {\rm few} \times m_\nu T^3$, and determine its fraction of the total
relativistic matter content of the universe. Relative to the energy density of
photons, $\rho_\gamma \sim T^4$, their fraction is
 \begin{equation} \label{eq:nufrac}
\Delta \omega_\nu = \frac{\rho_\nu}{\rho_{\gamma}} \agt 10
\times \frac{m_\nu}{T}  \, ,
 \end{equation}
where we are using that there are at least $4$ neutrino flavors
more massive than $m_\nu \sim 0.01 \, {\rm eV}$, and
that at least two are heavier than $3 m_\nu \sim 0.03 \, {\rm eV}$,
consistent with neutrino models as reviewed
in \cite{gelmini,whitenu}.
As the universe expands and $T$ decreases,
$\Delta \omega_\nu$ becomes larger, implying that the
massive neutrino species contribute progressively more
to the total relativistic energy density of the universe due
to the injection of energy from the swerves.
Thus as time goes on the relativistic fraction of the total energy density
of the universe {\it increases}. This is similar to what would
happen if there was some kind of late decay of nonrelativistic particles into
relativistic matter, suddenly repopulating the universe with relativistic,
hot dark matter.

The total amount of hot dark matter is bounded, implying that the
maximal value of $\Delta \Omega_\nu = \Delta \omega_\nu \, \Omega_\gamma$
is bounded too. At very late times, this gives the inequality
$\Delta \Omega_\nu < 0.2 \, \Omega_{DM}$  \cite{HDM,tzh}.
These bounds on $\Delta \Omega_\nu$ translate into constraints on the rate
of momentum diffusion induced by swerves. Indeed, suppose that the
swerve-induced momentum diffusion constant $k$ is large enough such
that the time scale $t_* \sim m^2_\nu/k$ after which nonrelativistic
neutrinos swerve back up to relativistic velocities is comparable to
the age of the universe where the neutrino would have frozen out
without swerves, $t_\nu \sim H^{-1}_\nu \sim (T_0/m_\nu)^{3/2}
H_0^{-1}$:
\begin{equation}
t_* \simeq t_\nu \, .
\end{equation}
If this were the case, neutrinos with mass $m_\nu$ would never
really cease to be relativistic, as we discussed above, since the
redshift from cosmic expansion would be compensated by the swerving
practically as soon as the temperature slides below $m_\nu$. From
this point onward, the neutrino species with masses $m_\nu$ would
contribute relativistic energy density that would scale as $1/a^3$.
Because $t_\nu \ll t_0$, today the
effective neutrino fraction of the relativistic
energy density would obey
\begin{equation} \label{eq:neutfr}
\Delta \omega_\nu > 10 \times \frac{m_\nu}{T_0} \, ,
\end{equation}
or, plugging in the value $T_0 \sim 10^{-4} \,
{\rm eV}$,
\begin{equation} \label{eq:neutfrnum}
\Delta \omega_\nu > 10^{5} \times \frac{m_\nu}{\rm
eV} \, ,
\end{equation}
which shows that for the neutrino masses $m_\nu > 0.01 \, {\rm eV}$, we would end up
with $\Delta \omega_\nu >  10^3$.
For larger $k$, $\Delta \omega_\nu$ would be even larger, due to continual energy buildup from slower swerving of relativistic particles.
This would translate to an estimate of neutrino energy density in the present universe
given by
\begin{equation}
\Delta \Omega_\nu = \Delta \omega_\nu \, \Omega_\gamma > 10^3 \, \Omega_\gamma \sim
 \Omega_{DM}
\, .
\end{equation}
which, being comparable with the cold dark matter density,
would violate the existing bounds \cite{HDM,tzh,wtz} by
almost an order of magnitude, in the very least. In this case
most of dark matter would be hot instead of cold.
This implies that in fact the heat up of the
neutrino gas by swerving should be reduced, until the time scale $t_*$
is slower than $t_\nu$. Requiring $t_* > t_\nu$ then yields
\begin{equation} \label{eq:kbound}
k < H_\nu m^2_\nu \simeq (\frac{m_\nu}{T_0})^{3/2} \, m^2_\nu \, H_0
\, .
\end{equation}
Substituting $m_\nu \sim 0.01 \, {\rm eV}$, we finally obtain the estimate
\begin{equation} \label{eq:numerical}
k < 10^{-61} \, {\rm GeV}^3 \, .
\end{equation}
This improves the bounds from hydrogen heating up of
\cite{DowkerSorkin} by about 17 orders of magnitude, and represents the
principal result of this work. We stress that this bound is somewhat
imprecise because we do not know the exact value of the neutrino
mass and the details of the balancing act between cosmic expansion
and swerving which sets the mean energy of the massive neutrino
relics. However, one would not expect parametrically large changes
of (\ref{eq:numerical}) from such corrections, as long as the
neutrino masses are not extremely small. We also stress that the
swerves controlled by a $k$ obeying (\ref{eq:numerical}) would only
impart minute effects on other particles. For example, a free
electron could only gain kinetic energy $\Delta E < 0.1 \, \mu {\rm
eV}$ throughout the age of the universe, clearly a miniscule amount
that would be easily masked away by thermal effects.

\subsection{Scale dependence of momentum space diffusion}

Now that we have constraints on $k$ it is interesting to rewrite
them in terms of our parameterization $k = c {\varepsilon}^{3-\gamma}
E_C^\gamma$.  The molecular masses that control the bounds for clouds,
hydrogen and nuclear stability are $m \sim {\rm few} \times
m_{proton}$ where $m_{proton} \sim 1 \, {\rm GeV}$ is the proton
mass, while the relevant mass scale for the cosmic neutrino bound is
$0.01 eV$. If we make the standard assumption that $E_C$, which
represents the scale of the causal set, is the Planck energy then we
can rewrite our parameterization as
\begin{equation}
\log c + \gamma (19- \log \varepsilon) = \log \frac {k} {\varepsilon^3} \, ,
\end{equation}
where all energies are measured in GeV. Our limits on $k$,
$k<10^{-61} \, {\rm GeV^3}$ at $\varepsilon = 0.01 \, {\rm eV}$ and
$k<10^{-44} \, {\rm GeV^3}$ at $\varepsilon = 1 \, {\rm GeV}$
then exclude a two-dimensional region in $\gamma, \log c$ parameter
space whose boundary
are the lines
\begin{eqnarray}
\log c + 19 \gamma = -44 \, \, {\rm for} \, \,  \gamma \leq \frac {16} {11} \, , \\
\log c + 30 \gamma = -28 \, \,{\rm for} \, \,  \gamma> \frac {16} {11} \, .
\end{eqnarray}

If we fix $\gamma=3$, as suggested by
EFT lore, then the corresponding constraint on $c$ is
$c< 10^{-118}$ which is an extremely small value begging for a first-principles explanation.
Presenting a detailed and {\it precise} explanation of
how $c$ could become so small is a challenge before causal set theory.
A possible way to
argue it may happen it is to demonstrate that there is a limit where swerving is completely
extinguished and so energy-momentum conservation and therefore translational symmetries are restored. In such an instance, one may believe that the coefficient $c$ could be much smaller
than unity even within the EFT description of swerving, if one existed, since the translational symmetries would then play a role of a softly broken symmetry \cite{gold}, protecting $c$. Alternatively, perhaps
causal set theory could never fully fit in the landscape of EFTs, in which case one must determine
whatever it is that emerges in the low energy limit and replaces EFT.

Thus one still faces the problem of actually calculating $k$ (or $c$) to determine precisely the link
between its absurdly small value and the broken translational symmetries.
This is reminiscent of the hierarchy problem in particle physics.
From the causal set perspective, one could try to link this to the observation that
if the sprinkling that models a random lattice
is denser, characterized by a larger value of
$E_C = V_C^{-1/4}$, then the probability of finding a
lattice point near a mass shell anywhere in spacetime increases, which might reduce swerving
and lower the value of $k$. In the EFT language, this would suggest that $k$ is {\it inversely} proportional to the `cutoff' $E_C$, which would imply $\gamma<0$
and effectively render $k$ very small, and swerving irrelevant, at low energies.
However such a detailed exploration of theoretical foundations of causal sets and the swerve effect
is beyond the scope of the present work.

~

\section{Summary}

In  this article we have considered low energy bounds on momentum
diffusion arising from swerves in causal set approach to quantum
gravity. We find that the bounds coming from old and cold
molecular clouds in our galaxy and nuclear stability are comparable
to the bounds on the diffusion coefficient $k$ coming from hydrogen
heating up of \cite{DowkerSorkin}. However, we find that the bounds
from the relic massive neutrino background coming from hot dark matter
constraints are much stronger, leading to our estimate $k < 10^{-61} \, {\rm
GeV}^3$ for neutrinos with masses $m_\nu > 0.01 \, {\rm eV}$, which would
improve the bounds of \cite{DowkerSorkin} by about 17 orders of
magnitude. We also note that if $\gamma=3$, which is strongly favored by
the standard effective field theory dimensional arguments, the
swerve-induced effects must be tremendously suppressed for some as
yet unknown reason. In this case the greatest effects of swerves would be on the light
cosmological neutrinos, whose relic energy density would show
surprising increase due to the energy injection from swerves. Oddly,
if our bound is almost saturated, such effects could lend to some
strange behavior in the equation of state of the universe. While
such possibilities are amusing, exploring them in detail must await
further developments in causal set theory.

\acknowledgments

We would like to thank M. Bucher, H.F. Dowker, L. Knox, K.A. Olive, R.
Sorkin and especially M. Kaplinghat for useful discussions.
We would also like to thank M. Kaplinghat for comments on the manuscript.
NK would like to thank Galileo
Galilei Institute, Florence, Italy, and to LPT, Universite de
Paris-Sud, Orsay, France, for kind hospitality during the course of
this work. NK and DM were supported in part by the DOE Grant
DE-FG03-91ER40674. NK was also supported in part by the NSF Grant
PHY-0332258 and by a Research Innovation Award from the
Research Corporation.


\begin{thebibliography}{99}
\bibitem{broad}
Swerving may still lead to the broadening of spectral lines
of particles that hop about the lightcone, as argued in  \cite{DowkerSorkin}.
\bibitem{gold}
{If so, then there may be Goldstone bosons
associated with the broken translational symmetries, which
may be very light and hence could mediate new long range
forces. Exploring their dynamics could strengthen the bounds
on translational symmetry breaking.}
\end{thebibliography}
\end{document}